\documentclass[12pt]{article}  

\RequirePackage{graphicx,color}
\usepackage{amssymb,amsmath,slashed,mathrsfs,cite,hyperref}

\usepackage{bbold,parskip}
\newcommand{\gdualn}[1]{\overset{\:{}^{{}^{\boldsymbol{\neg}}}}{\smash[t]{#1}}} 
\newcommand{\0}{\mbox{\boldmath$\displaystyle\mathbb{O}$}}

\newcommand{\J}{\mbox{\boldmath$\displaystyle\boldsymbol{J}$}}

\def\s{\mbox{\boldmath$\displaystyle\boldsymbol{\sigma}$}}
\def\p{\mbox{\boldmath$\displaystyle\boldsymbol{p}$}}
\def\0{\mbox{\boldmath$\displaystyle\boldsymbol{0}$}}

\begin{document}

\date{\empty}

\noindent
\textbf{\large A critical response to Moshe  Chaichian and colleagues on mass dimension one fields, and related matters }

\vspace{11pt}

\noindent
{\textbf{Dharam Vir Ahluwalia}}$^{a,1}$,
\textbf{Cheng-Yang Lee}$^{b,2}$
\vspace{11pt}
\textcolor{red}{\hrule}
\vspace{11pt}
\begin{quote}
\small{
\noindent $^a$ Center for the Studies of the Glass Bead Game, Notting Hill,\\  Victoria 3168, Australia   \\ 
        $^b$   Center for Theoretical Physics, College of Physics, Sichuan University, Chengdu, 610064, China } \\ \\
$^1$ Email: dharam.v.ahluwalia@gmail.com 
 \\ ORCID: \href{https://orcid.org/0000-0002-6331-6243}{https://orcid.org/0000-0002-6331-6243}, Corresponding author\\
$^2$ E-mail:  cylee@scu.edu.cn \\
ORCID: \href{https://orcid.org/0000-0002-6926-2717}{https://orcid.org/0000-0002-6926-2717}
\end{quote}

\vspace{11pt}
\begin{quote}
\textbf{Abstract.}
In a recent article published in EPJC, Moshe Chaichian and colleagues have presented an overarching critique of mass dimension one fields and related matters. In this Letter we show without ambiguity where these authors go wrong. 

 \vspace{11pt}

\noindent
\textbf{Keywords. } {Elko, Mass dimension one fermions, spin-half bosons, Wigner degeneracy, dark matter, dark energy}

\end{quote}
\newpage
\vspace{11pt}
\noindent

Authors of reference~\cite{Aguirre:2021whp} claim that the 
the quantum field theoretic formalism of 
mass dimension one fields~\cite{AHLUWALIA20221,ahluwalia_2019,Ahluwalia_2022} is not physically viable on the grounds that the formalism cannot satisfy the rotational constraint. They make the same claim in the context of mass dimension three half bosons, and assert that duals and adjoints of expansion coefficients and the associated fields do not enjoy the freedom we invoke. We here give briefest of arguments to show that these authors are completely mistaken in all their claims.


To see the necessity of a new spinorial dual 
let $\lambda(\boldsymbol{\p})$ be a Majorana spinor associated with mass $m\ne 0$. Its right-transforming component has opposite spin projection when compared with its left-transforming component. So it is genuinely different from a Dirac spinor.  Furthermore, if we define its dual as
\begin{equation}
\overline{\lambda}(\p) \stackrel{\mathrm{def}}{=} 
\lambda^\dagger(\boldsymbol{\p}) \gamma_0
\end{equation}
as is often done in the literature
then $\overline{\lambda}(\p) \lambda(\p)$  identically vanishes~\cite{Aitchison:2004cs}. Unless one deviates from the Dirac formalism and promotes the Majorana spinors to Grassmann variables, the mass term identically vanishes and for that reason one cannot construct a Lagrangian density~\cite{Aitchison:2004cs}. In the just cited reference, this is used as a justification to promote Majorana spinors to Grassmann variables.
A well prepared reader knows that avoiding this step led in 2004 to the ``unexpected theoretical discovery'' of mass dimension one fields. The line of arguments leading to this conclusion is well documented~\cite{AHLUWALIA20221}.

To proceed further, consider a $\lambda(\p)$ with its left transforming component with  spin projection  $+1/2$. Call this spinor as $\lambda_+(\p)$.  Next 
consider a $\lambda(\p)$ with its left transforming component with  spin projection $-1/2$. Call this spinor as $\lambda_-(\p)$.  Then, with $\mathcal{D} \stackrel{\mathrm{def}}{=} m^{-1} \gamma_\mu p^\mu$ one readily finds that Majorana spinors can not be  eigenspinors of 
$\mathcal{D}$, but instead, modulo factors of $\pm i$, $\mathcal{D}$ takes $ \lambda_+(\p)$ into  $\lambda_-(\p)$ and vice versa
\begin{equation}
\mathcal{D} \lambda_+(\p) = i  \lambda_-(\p), \quad
\mathcal{D} \lambda_-(\p) = - i  \lambda_+(\p), \label{eq:dvo}
\end{equation} 
To the best of our knowledge, result (\ref{eq:dvo}) was first obtained by 
Valeri Dvoeglazov~\cite{Dvoeglazov:1995eg}, and only later refined in~\cite{ahluwalia_2019}. 

This readies the stage to define a new dual defined as 
\begin{equation}
\gdualn{\lambda}(\p) \stackrel{\mathrm{def}}{=}\left[ \mathcal{D}  \lambda(\p)\right]^\dagger \gamma_0. \label{eq:lambdadual}
\end{equation}
Under this dual, $\gdualn{\lambda}(\p) \lambda(\p)$ 
yields a non-zero norm. All this is well known to the community of people working on the subject. Our only reason to repeat this in this briefest of reviews is to show how duals, and the associated adjoints, come to be, and to point out how claims of reference~\cite{Aguirre:2021whp} are totally in error.

We now make a transition from Majorana spinors to Elko. The charge conjugation operator 
\begin{equation}
\mathcal{C} = 
\left(
\begin{array}{cc}
\0 & i \Theta\\
-i\Theta &\0
\end{array}
\right) K.
\end{equation}
where $K$ complex conjugates to the right,
has two doubly degenerate eigenvalues.\footnote{Because of the nonlinearity of $\mathcal{C}$ we have a choice of choosing these to be real, or complex. We choose these to be real.}
Two of these account for Majorana spinors, and we take it to be $+1$. The other two, which we choose to be $-1$,  are associated with  the anti self conjugate spinors. Together, these form what in literature are called Elko and are represented by self conjugate spinors
$\lambda_\pm^S(\p)$, and anti self conjugate spinors, 
$\lambda_\pm^A(\p)$.

 Because the charge conjugation operator is \emph{an antilinear} operator, the product, $ i \times \lambda_\pm^A(\p)$ is  self conjugate under $\mathcal{C}$. Similarly,  the product $ i \times \lambda_\pm^S(\p)$ yields an anti self conjugate spinor. We thus have four self conjugate Elko, and four anti self conjugate Elko. This doubling in the degrees of freedom is over and above that which arises from the two fold degeneracy under $\mathcal{C}$. Such a possibility  was first envisioned by Wigner~\cite{Wigner:1962ep}.

These can be used as expansion coefficients of a quantum field that carries a two fold Wigner degeneracy. Such  a spin one half field has mass dimension one and can satisfy fermionic, and surprisingly also the bosonic statistics. 

Our remaining task is two fold.  First, we 
must argue as to how the rotational symmetry is preserved in our formalism. Second, we must show how spin one half bosons arise in the mass dimension three half, as well mass dimension one case.



Generally, it is assumed that the expansion coefficients of a quantum field at rest are completely unconstrained as long as they form a complete set. A detailed study of Weinberg's  quantum theory of fields~\cite{Weinberg:1995mt} indeed finds this widespread misconception 
needs to be corrected.
 The authors of reference~\cite{Aguirre:2021whp} indeed pay a superficial attention to his analysis. In fact,  their ``analysis'' verges on plagiarism of Weinberg's profound work.
Furthermore, they apply their analysis to the wrong degrees of freedom! In this instance, the correct field to apply the rotational constraints must have a two fold Wigner degeneracy. When this is done, one indeed finds that the said rotational constraints are fully satisfied.

Here is the brief  outline of  how the calculation on rotational symmetry proceeds. The rotational constraints read
   \begin{equation}
 \sum_{{\sigma}^\prime} \lambda^S_{{\ell^\prime}}(\0,\sigma^\prime)\, \J_{{\sigma}^\prime\sigma} = \sum_\ell  \boldsymbol{\mathcal{J}}_{{\ell}^\prime\ell} \,\lambda^S_\ell(\0,\sigma),\label{eq:rotp}
 \end{equation}
 and 
  \begin{equation}
 \sum_{{\sigma^\prime}} \lambda^A_{{\ell^\prime}}(\0,\sigma^\prime) \,\J^\ast_{{\sigma^\prime}\sigma} = - \sum_\ell \boldsymbol{\mathcal{J}}_{{\ell^\prime}\ell}\, \lambda^A_\ell(\0,\sigma).\label{eq:rota}
 \end{equation}
The $\lambda^S(\0,\sigma)$ and $\lambda^A(\0,\sigma)$ are the expansion coefficients at rest of the quantum field with a two fold Wigner degeneracy, and thus $\sigma=1,2,3,4$.  The $\boldsymbol{\mathcal{J}}$
are
 \begin{align}
 & \mathcal{J}_x =  \frac{1}{2} \left(
 \begin{array}{cc}
 \sigma_x & \mathbb{o} \\
 \mathbb{o} & \sigma_x
 \end{array}
 \right),\quad \mathcal{J}_y =  \frac{1}{2} \left(
 \begin{array}{cc}
 \sigma_y  & \mathbb{o} \\
 \mathbb{o} & \sigma_y
 \end{array}
 \right),\nonumber  \\ 
 &\mathcal{J}_z =  \frac{1}{2} \left(
 \begin{array}{cc}
 \sigma_z & \mathbb{o} \\
 \mathbb{o} & \sigma_z
 \end{array}
 \right).\label{eq:CurlyJ14}
  \end{align}
 where $\s$ are the usual Pauli matrices with $\sigma_z$ diagonal.   If the formalism is to satisfy (\ref{eq:rotp}) and (\ref{eq:rota}),
 there must exist an $\mathfrak{su}(2)$ satisfying $4\times 4$ matrices $\J$ with \emph{doubly} degenerate eigenvalues $\pm1/2$. Such a $\J$ exists with the consequence that the formalism of mass dimension one fields satisfies the rotational constraints. The calculational difference with the Weinberg's formalism \emph{without} Wigner degeneracy for spin one half is that Weinberg takes $2\times 2$ matrices $\J$ as input and obtains the particle and antiparticle expansion coefficients as an output.   Here, apart from matters arising from Wigner degeneracy, the calculations feed into the rotational constraints the particle and antiparticle expansion coefficients at rest and solve for the $4\times 4$ $\J$ matrices.
 
Finally, to attend to the last concern of  Moshe Chaichian and colleagues, we note that generations of physicists have learned that the quantisation of the Dirac field necessarily requires the use of anti-commutators for the field-field (and also for the conjugate momentum) at space-like separation for the Hamiltonian to be positive definite. A nuanced exposition of this is not required as anyone reading this critique knows this as an irrefutable wisdom. There is however a caveat, an implicit assumption,  that duals of the Dirac spinors must necessarily be $\psi^\dagger(\p) \gamma_0$, and not be in parallel with (\ref{eq:lambdadual}): $\gdualn{\psi}(\p) \stackrel{\mathrm{def}}{=}\left[ \mathcal{D}  \psi(\p)\right]^\dagger \gamma_0$.  The use of such a dual changes the sign of  the norm of the antiparticle spinors. When this sign is propagated through quantum field theoretic calculations, one finds that the Dirac fields have positive definite hamiltonians, and are local. The statistics: bosonic~\cite{Ahluwalia_2022}.\footnote{Similar remarks apply to the formalism of mass dimension one bosons.}

For the reasons outlined here we categorically assert that the paper of Moshe Chaichian and colleagues is based on deep rooted misconceptions, and application of a profound formalism of Weinberg to a wrong quantum field. More details shall be published elsewhere.

\newpage
\noindent
\textbf{Acknowledgements}

DVA is supported by the Center for the Studies of the Glass Bead Game, 
CYL is supported by the  Sichuan University Postdoctoral Research Fund No.2022SCU12119


\providecommand{\href}[2]{#2}\begingroup\raggedright\endgroup

\end{document}